\documentclass[11pt,a4paper,english,superscriptaddress,aps,preprint]{revtex4}
\usepackage[T1]{fontenc}
\usepackage[latin9]{inputenc}
\usepackage{amsmath}
\usepackage{amssymb}
\usepackage{esint}

\makeatletter

\makeatletter\usepackage{babel}

\usepackage{hyperref}

\makeatother

\begin{document}

\title{Ho$\breve{{\rm r}}$ava-Lifshitz modifications of the Casimir effect}

\author{A. F. Ferrari}
\affiliation{Centro de Ci\^encias Naturais e Humanas, Universidade Federal do
ABC\\Rua Santa Ad\'elia, 166, 09210-170, Santo Andr\'e, SP, Brazil}
\email{alysson.ferrari@ufabc.edu.br}

\author{H. O. Girotti}
\affiliation{Instituto de F\'{\i}sica, Universidade Federal do Rio Grande do
Sul\\Caixa Postal 15051, 91501-970, Porto Alegre, RS, Brazil}
\email{hgirotti@if.ufrgs.br}

\author{M. Gomes}
\affiliation{Instituto de F\'{\i}sica, Universidade de S\~ao Paulo\\Caixa Postal 66318, 05315-970, S\~ao Paulo, SP, Brazil}
\email{mgomes,ajsilva@fma.if.usp.br}

\author{A. Yu. Petrov}
\affiliation{Departamento de F\'{\i}sica, Universidade Federal da Paraíba\\Caixa Postal 5008, 58051-970, Jo\~ao Pessoa, Para\'{\i}ba, Brazil}
\email{petrov@fisica.ufpb.br}

\author{A. J. da Silva}
\affiliation{Instituto de Física, Universidade de São Paulo\\Caixa Postal 66318, 05315-970, São Paulo, SP, Brazil}
\email{mgomes,ajsilva@fma.if.usp.br}

\begin{abstract}
We study the modifications induced by spacetime anisotropy on the
Casimir effect in the case of two parallel plates. Nonperturbative
and perturbative regimes are analyzed. In the first case the Casimir
force either vanishes or it reverses its direction which, in any case,
makes the proposal untenable. On the other hand, the perturbative
model enables us to incorporate appropriately the effects of spacetime
anisotropy.
\end{abstract}

\maketitle

\section{Introduction}

A great deal of attention has been devoted to the Ho$\breve{{\rm r}}$ava-Lifshitz
(HL) theory\,\cite{Hor} since it might give rise to the existence
of a renormalizable formulation of quantum gravity. The essence of
the Ho$\breve{{\rm r}}$ava proposal consists in attributing different
scaling properties to space and time coordinates: $x^{i}\to bx^{i}$,
$t\to b^{z}t$, where $z$ is a critical exponent characterizing the
ultraviolet behavior of the theory. Power counting suffices to suggest
the renormalizability of four-dimensional quantum gravity for $z=3$.
One expects to recover the Lorentz symmetry at the infrared limit.
Different issues related to the HL gravity, including its cosmological
aspects\,\cite{Hor1,HorCos,ed1}, exact solutions\,\cite{Lu}, and
black holes\,\cite{BH} have already been presented in the literature.

We dedicate this work to study the consequences of the spacetime anisotropy
on the Casimir effect for two parallel plates\,\cite{Casimir}. Other
consequences of the HL proposal in field and brane theories have been
considered in\,\cite{ed1,ed2,Kluson,Anselmi,cpn,ff}. We consider
two different models. One is essentially nonperturbative and therefore
closer in spirit to the original HL proposal. It is contained in Section
II. The second one is tailored in such a way that the spacetime anisotropy
is a correction to the standard Casimir effect. This is our Section
III. Section IV contains the conclusions.

\section{The nonperturbative model}

The model is described by the following action (repeated Latin indices
sum from one to $d$) \begin{equation}
S=\frac{1}{2}\int dtd^{d}x\left(\partial_{0}\phi\partial_{0}\phi-\ell^{2\left(z-1\right)}\partial_{i_{1}}\partial_{i_{2}}\ldots\partial_{i_{z}}\phi\partial_{i_{1}}\partial_{i_{2}}\ldots\partial_{i_{z}}\phi\right)\,.\label{eq:1}\end{equation}
 Notice that the number of spatial derivatives acting on the real
scalar field $\phi$ is equal to $2z$, where $z$ is the above mentioned
critical exponent. As required by consistency, the constant $\ell$
has dimension of length. 

As in Ref.\,\cite{Casimir}, we start by considering two large parallel
plates with area $L^{2}$, with a separation $a$ ($a\ll L$). Let
these plates be orthogonal to the $x_{3}$ axis. The free scalar field
$\phi$, being the solution of the equation of motion \begin{equation}
\left(\partial_{0}^{2}-\ell^{2\left(z-1\right)}\left(-1\right)^{z}\Delta^{z}\right)\phi=0\,,\label{eq:2}\end{equation}
 is localized between these plates, and verifies, by assumption, Dirichlet
boundary conditions, i.e., $\phi(x_{3}=0)=\phi(x_{3}=a)=0$. The solving
of this equation leads to \begin{equation}
\phi=\sqrt{\frac{2}{a}}\,\frac{1}{2\pi}\,\sum\limits _{n=0}^{\infty}\int\frac{d^{2}k}{2\omega_{\vec{k},n}}\,\sin\left(\frac{\pi nx_{3}}{a}\right)\left(a_{n}\left(k\right)e^{ikx}+a_{n}^{\dagger}\left(k\right)e^{-ikx}\right)\,,\label{eq:3}\end{equation}
 where $n$ is an integer, $kx\equiv k_{0}x_{0}-k_{1}x_{1}-k_{2}x_{2}$,
\begin{equation}
k_{0}=\ell^{z-1}\omega_{\vec{k},n}^{z}\,,\label{eq:4}\end{equation}
 and\begin{equation}
\omega_{\vec{k},n}=\sqrt{k_{1}^{2}+k_{2}^{2}+\left(\frac{\pi n}{a}\right)^{2}}\,.\label{eq:5}\end{equation}

The Hamiltonian operator emerging from canonical quantization reads
\begin{equation}
H=\frac{\ell^{z-1}}{2}\sum\limits _{n=1}^{\infty}\int\frac{d^{2}k}{2\omega_{\vec{k},n}}\,\left(a_{n}a_{n}^{\dagger}+a_{n}^{\dagger}a_{n}\right)\omega_{\vec{k},n}^{z},\label{eq:6}\end{equation}
where $a_{n}(\vec{k})$ and $a_{n}^{\dagger}(\vec{k})$ are, respectively,
annihilation and creation operators obeying the commutation relation
\begin{equation}
\left[a_{n}(\vec{k}),a_{n^{\prime}}^{\dagger}(\vec{k}^{\prime})\right]=2\omega_{\vec{k},n}\delta_{nn^{\prime}}\delta^{2}(\vec{k}-\vec{k}^{\prime})\,.\label{eq:7}\end{equation}
 Hence, the vacuum energy is given by \begin{equation}
E=\left\langle 0|H|0\right\rangle =\frac{\ell^{z-1}L^{2}}{8\pi^{2}}\sum\limits _{n=1}^{\infty}\int d^{2}k\,\omega_{\vec{k},n}^{z}.\label{eq:8}\end{equation}

This last integral is divergent and its evaluation demands the introduction
of a regularization procedure. In particular, we choose to replace
Eq.\,(\ref{eq:8}) by \begin{equation}
E_{\mbox{reg}}=\frac{\ell^{z-1}L^{2}}{8\pi^{2}}\sum\limits _{n=1}^{\infty}\int d^{2}k\,\omega_{\vec{k},n}^{z}e^{-\alpha\omega_{\vec{k},n}},\label{eq:9}\end{equation}
where $\alpha$ is a semipositive real parameter to be set to zero
at the end of the calculations. By performing the angular integration,
one finds \begin{equation}
E_{\mbox{reg}}=\frac{\ell^{z-1}L^{2}}{4\pi}\sum\limits _{n=1}^{\infty}\int_{\frac{\pi n}{a}}^{\infty}d\omega_{\vec{k},n}\,\omega_{\vec{k},n}^{z+1}e^{-\alpha\omega_{n}}\,,\label{eq:11}\end{equation}
 so that \begin{align}
E_{\mbox{reg}} & =\frac{\ell^{z-1}L^{2}}{4\pi^{2}}\left(-1\right)^{z+1}\left(\frac{\partial}{\partial\alpha}\right)^{z+1}\left[\frac{a}{\alpha^{2}}\sum\limits _{n=0}^{\infty}\frac{B_{n}}{n!}\left(\frac{\pi\alpha}{a}\right)^{n}\right]\,,\label{eq:12}\end{align}
 where the $B_{n}$ are the Bernoulli numbers.

We shall next subtract from this last expression the vacuum energy
in the absence of plates, namely,\begin{align}
E_{\mbox{reg}}^{0} & =\frac{\ell^{z-1}L^{2}}{4\pi^{2}}\left(-1\right)^{z+1}\left(\frac{\partial}{\partial\alpha}\right)^{z+1}\frac{a}{\alpha^{2}}\,.\label{eq:13}\end{align}
 Thus, \begin{align}
E_{\mbox{reg}}-E_{\mbox{reg}}^{0} & =\frac{\ell^{z-1}L^{2}}{4\pi^{2}}\left(-1\right)^{z+1}\left(\frac{\partial}{\partial\alpha}\right)^{z+1}\left[\frac{a}{\alpha^{2}}\sum\limits _{n=1}^{\infty}\frac{B_{n}}{n!}\left(\frac{\pi\alpha}{a}\right)^{n}\right]\,.\label{eq:14}\end{align}
 However, as $\alpha\to0$ only the term $n=z+3$,\begin{equation}
\frac{\ell^{z-1}L^{2}}{4\pi^{2}}\left(-1\right)^{z+1}\,\frac{\pi^{z+3}B_{z+3}}{\left(z+3\right)\left(z+2\right)}\,\frac{1}{a^{z+2}}\,,\label{eq:15}\end{equation}
 contributes to the modified Casimir force\begin{equation}
F=-\frac{d}{da}\lim_{\alpha\rightarrow0}\left(E_{\mbox{reg}}-E_{\mbox{reg}}^{0}\right)\,.\label{eq:15a}\end{equation}

The resulting force per unit area turns out to be\begin{equation}
\frac{F_{z}}{L^{2}}=\left(-1\right)^{z-1}\,\frac{\ell^{z-1}\pi^{z+1}B_{z+3}}{4\left(z+3\right)}\,\frac{1}{a^{z+3}}\,.\label{eq:16}\end{equation}
 For the standard case, $z=1$, one recovers the well-known result
\begin{equation}
\frac{F_{1}}{L^{2}}=-\frac{\pi^{2}}{480\, a^{4}}\,,\label{eq:17}\end{equation}
 since $B_{4}=-1/30$.

For other values of $z$, we found instructive to compute the relative
intensity of the Casimir force referred to that of the standard case,
i.e.,\begin{equation}
\frac{F_{z}}{F_{1}}=\left(-1\right)^{z}\times120\frac{\pi^{z-1}B_{z+3}}{\left(3+z\right)}\left|\frac{\ell}{a}\right|^{z-1}\,.\label{eq:19}\end{equation}
The relevant points here are that $F_{2}$ vanishes while the direction
of $F_{3}$ is opposite to that of $F_{1}$. This indicates that the
model defined in Eq. (\ref{eq:1}) is untenable.

\section{The perturbative model}

We now propose the action \begin{equation}
S^{\prime}=\frac{1}{2}\int dtd^{d}x\,\left(\partial_{0}\phi\partial_{0}\phi-\partial_{i}\phi\partial_{i}\phi-\ell^{2\left(z-1\right)}\partial_{i_{1}}\partial_{i_{2}}\ldots\partial_{i_{z}}\phi\partial_{i_{1}}\partial_{i_{2}}\ldots\partial_{i_{z}}\phi\right)\,,\label{eq:20}\end{equation}
where the spacetime anisotropy shows up as a modification of the standard
free Lagrangian. The dispersion relation modifies as follows,\begin{align}
k_{0} & =\omega_{\vec{k},n}\sqrt{1+\left(\ell\omega_{\vec{k},n}\right)^{2\left(z-1\right)}}\,,\label{eq:21}\end{align}
 where $\omega_{\vec{k},n}$ was already defined in Eq. (\ref{eq:5}).
The vacuum energy reads\begin{equation}
E=\frac{L^{2}}{8\pi}\sum\limits _{n=1}^{\infty}\int_{\frac{n\pi}{a}}^{\infty}d\omega_{\vec{k},n}\,\omega_{\vec{k},n}^{2}\sqrt{1+\left(\ell\omega_{\vec{k},n}\right)^{2\left(z-1\right)}}\,.\label{eq:22}\end{equation}
 By adopting the same regularization as in Section II, \begin{equation}
E_{\mbox{reg}}=\frac{L^{2}}{8\pi}\sum\limits _{n=1}^{\infty}\int_{\frac{n\pi}{a}}^{\infty}d\omega_{\vec{k},n}\,\omega_{\vec{k},n}^{2}\sqrt{1+\left(\ell\omega_{\vec{k},n}\right)^{2\left(z-1\right)}}e^{-\alpha\omega_{\vec{k},n}}\,,\label{eq:23}\end{equation}
 we simplify the integrand by using that \begin{align}
\sqrt{1+\left(\ell\omega_{\vec{k},n}\right)^{2\left(z-1\right)}} & \sim1+\frac{1}{2}\left(\ell\omega_{\vec{k},n}\right)^{2\left(z-1\right)}\,.\label{eq:24}\end{align}
 Then, the vacuum energy turns out to be given by the sum of the usual
one plus the correction\begin{equation}
\delta E_{\mbox{reg}}=\frac{L^{2}\ell^{2\left(z-1\right)}}{16\pi}\sum\limits _{n=1}^{\infty}\left(-\frac{\partial}{\partial\alpha}\right)^{2z}\int_{\frac{n\pi}{a}}^{\infty}d\omega_{\vec{k},n}\, e^{-\alpha\omega_{\vec{k},n}}\,.\label{eq:25}\end{equation}
 This expression can be cast \begin{align}
\delta E_{\mbox{reg}} & =\frac{L^{2}\ell^{2\left(z-1\right)}}{16\pi^{2}}\left(\frac{\partial}{\partial\alpha}\right)^{2z}\left[\frac{a}{\alpha^{2}}\sum\limits _{n=0}^{\infty}\frac{B_{n}}{n!}\left(\frac{\pi\alpha}{a}\right)^{n}\right]\,.\label{eq:27}\end{align}
One is next to subtract the vacuum energy in the absence of the plates,
which corresponds to the $n=0$ term in the sum above. Moreover, only
the term $n=2z+2$,\begin{equation}
\frac{L^{2}}{16\pi^{2}}\,\frac{\ell^{2\left(z-1\right)}\pi^{2z+2}B_{2z+2}}{\left(2z+2\right)\left(2z+1\right)a^{2z+1}}\,,\label{eq:28}\end{equation}
 happens to contribute to the Casimir force (\ref{eq:15a}).

Thus, the total resulting force per unit area is\begin{equation}
\frac{F+\delta F}{L^{2}}=-\frac{\pi^{2}}{480\, a^{4}}\left[1-30\,\frac{\pi^{2z-2}B_{2z+2}}{\left(2z+2\right)}\,\left(\frac{\ell}{a}\right)^{2\left(z-1\right)}\right]\,.\label{eq:29}\end{equation}
We emphasize that the correction induced by the spacetime anisotropy
never vanishes. However, to make sense, the ratio $\ell/a$ must be
small enough so as to secure that the just mentioned correction remains
below experimental uncertainties. To be more precise, for an experimental
uncertainty of order 1\%\,\cite{lamoreaux,mohideen} we estimate
$\left|\ell/a\right|\lesssim0.17$ for $z=3$ and $\left|\ell/a\right|\lesssim0.09$
for $z=2$. In\,\cite{lamoreaux,mohideen}, the separation $a$ is
as small as $10^{-7}\mbox{m}$, so in principle $\ell$ could be at
the most of order $10^{-8}\mbox{m}$.

\section{Conclusions}

We have shown that the Casimir force is indeed modified by the presence
of spacetime anisotropy. The effect may be drastic, up to the point
of changing the direction of the force, which cannot be accommodated
within current experimental observations. Alternatively, one may implement
the anisotropy in such a way that the effects it produces remain acceptable
from the experimental point of view.

\textbf{\vspace{1cm}
 }

\textbf{Acknowledgments. }This work was partially supported by the
Brazilian agencies Conselho Nacional de Desenvolvimento Cient\'{\i}fico
e Tecnol\'{o}gico (CNPq), Funda\c{c}\~{a}o de Amparo \`{a} Pesquisa
do Estado de S\~{a}o Paulo (FAPESP), and Coordena\c{c}\~{a}o de
Aperfei\c{c}oamento de Pessoal de N\'{\i}vel Superior (CAPES).

\end{document}